\documentclass[12pt]{article}
\usepackage[affil-it]{authblk}
\usepackage{graphicx}
\usepackage{url}
\usepackage{pstricks}
\usepackage{color}
\usepackage{epsfig}
\usepackage{fancyhdr}
\usepackage{mathbbol}
\usepackage{dsfont}
\usepackage{pstricks}
\usepackage{bbding}
\usepackage{color}
\usepackage{bbm}
\usepackage{pifont}
\usepackage{bm}
\usepackage{multirow}
\usepackage{bm}
\usepackage{amsmath}
\usepackage{amsfonts}
\usepackage{amssymb}

\def\be{\begin{equation}}
\def\ee{\end{equation}}
\def\bea{\begin{eqnarray}}
\def\eea{\end{eqnarray}}

\title{The $Z(4430)$ and a New Paradigm for Spin Interactions in Tetraquarks}
\author{L. Maiani$^{*,+}$, F. Piccinini$^\dag$, A. D. Polosa$^*$ and V. Riquer$^{*,+}$
}
\affil{\small \small $^*$Dipartimento di Fisica, Sapienza Universit\`a di Roma, and INFN, Sezione di Roma, Piazzale Aldo Moro 2, I-00185 Roma, Italy\\ $^\dag$INFN, Sezione di Pavia, Via A. Bassi 6, I-27100, Pavia, Italy\\
$^+$ Instituto de Fisica Teorica, IFT-UAM/CSIC, Universidad Autonoma de Madrid, Cantoblanco, 28049, Madrid, Spain
}
\begin{document}

\maketitle
\begin{abstract}
\noindent
Following the recent confirmation of the  $Z^+(4430)$ resonance with $J^{PG}=1^{++}$, we have re-examined the model of $S$ and $P$ wave tetra- -quarks. We propose a `type-II'  diquark-antidiquark model  which shows to be very effective at producing a simple and comprehensive picture of the  $J^{PG}=1^{++}$ and $1^{--}$ sectors of the recently discovered charged tetraquarks  and of the observed $Y$ resonances. The model is still faced with the unresolved difficulty of explaining why some states seem to have incomplete isospin multiplets. 
\end{abstract}

\subsection*{Introduction}
The recent observation by LHCb~\cite{z4430lhcb} of the $Z(4430)$ charged tetraquark supports the earlier BELLE~\cite{z4430belle,z4430belle2} indication,  later cast into doubt by BaBar~\cite{Aubert:2008aa}. This confirmation 
is decisive for the understanding of the complex system of charmonium-like exotic resonances, the so called $X,Y,Z$ states, discovered in recent years and the subject of diverse and conflicting theoretical interpretations. 

The idea that  charmed meson molecules might be formed in high energy reactions, proposed originally in~\cite{De Rujula:1976qd}, has been invoked several times in the context of $X,Y,Z$ spectroscopy with special reference to loosely-bound molecules whose prototypical example is the $X(3872)$, supposed to be a $D^0\bar D^{*0}$ molecule with a binding energy ${\cal E}\approx 0$ and a width $\Gamma\lesssim 1$~MeV. Such an idea appears to be  at odds with the large prompt production cross sections observed at CDF~\cite{cdfprompt} and CMS~\cite{cmsprompt}, as is confirmed by the calculations done with hadronization algorithms~\cite{bigna}. Final state interactions within the $D^0\bar D^{*0}$ pair are often invoked as effective in coalescing the pair into a barely bound state, even if the components are initially recoiling with high relative momenta~\cite{arto}. The limits of such an approach are further discussed in~\cite{inprep}. 

The molecular picture has also been proposed  to explain the nature of the $Z(4430)$ resonance. In this case, however, the loosely bound mechanism does not work as there are no open charm thresholds with $J^{PG}=1^{++}$ quantum numbers at that mass.
In~\cite{Rosner} it is suggested that the $Z(4430)$ might be a $D^*(2010)\bar D_1(2420)$ bound state in $S-$wave, but this has $J^P=0^-,1^-,2^-$, not consistent with the recent observations strongly suggesting $J^P=1^+$.  
For the  molecular picture see also~\cite{Meng}, \cite{Wang:2013cya}. Other theoretical interpretations include $\Lambda_c \bar{\Sigma}_c$ baryonium ~\cite{Qiao},  cusp effect~ \cite{Bugg}, $D_s$ radial excitation~\cite{Matsuki}, as well as sum rules calculations  based on the $D^{*}\bar{D}_1$ molecule~\cite{Lee},\cite{Bracco}. All these speculations envisage effects due to the residual, short range, forces generated by colorless meson exchange between color neutral objects. 

Here we shall follow the tetraquark interpretation of states made by {\it colored components}, diquarks and antidiquarks, bound by the long range color forces~\cite{xmai}. 
Hidden beauty tetraquark states have been considered in~\cite{aali}.

A diquark is made by a $[cq]$ pair in a color antisymmetric state, with $c$ the charm and $q$ a light, $u$ or $d$, quark. 
This picture supports the existence of bound states with higher orbital angular momentum and/or  radially excited and is consistent with production at Tevatron and LHC with cross sections similar to the ones of normal charmonia. 

In a 2007 paper~\cite{z4430}, after the observation of the  charged $Z(4430)$ resonance by Belle, and judging from the considerably higher mass with respect to lowest lying tetraquark, we proposed this resonance to be the first radial excitation of a $1S$ state. The $Z(4430)-X(3872)$ mass difference is indeed very similar to the $\psi(2S)-\psi(1S)$ mass difference and in line with the observed decay
\begin{equation}
Z^+ (4430)\to \pi^+ + \psi(2S)
\label{radexc}
\end{equation}
We also noted: {\it `A crucial consequence of a $Z(4430)$ charged particle is that a charged state decaying into $J/\psi +\pi^{\pm}$ or $\eta_c + \rho^{\pm}$ should be found around 3880 MeV'}. 

Now that the $Z_c(3900)$ has been seen by BES III~\cite{Ablikim:2013mio} and Belle~\cite{Liu:2013dau} with the decay
\begin{equation}
Z^+ (3900)\to \pi^+ + \psi(1S)
\end{equation}
and a neutral partner suggested by CLEO~\cite{Xiao:2013iha}, 
and with the further observation of $Z(4020)$ by the BES III  Collaboration~\cite{4020},\cite{Ablikim:2013emm}, the tetraquark picture looks more attractive and constrained as compared to some years ago~\cite{noirassegna}.

With only the $X(3872)$ at hand, the couplings characterizing spin interactions of the different flavors were deduced in~\cite{xmai} from the spectrum  of mesons and baryons under rather uncontrolled hypotheses, such as the one-gluon exchange approximation and the equality of $|\psi(0)|^2$, the overlap probability of quarks and antiquarks in mesons or baryons and in the tetraquark,  see~\cite{maiaerice}.

We introduce in this paper a `type-II' model, based on a simple, new Ansatz on spin-spin couplings, whereby {\it the $cq$ interaction inside the diquark is assumed to dominate} over all other possible pairings. A value of this coupling: $\kappa_{qc}\approx 67$~MeV, larger than the one deduced in~\cite{xmai} from baryon masses, explains the near degeneracy of $X(3872)$ with $Z(3900)$ as well as the $Z(4020)-Z(3900)$ mass difference. Predictions for the other $S$ wave tetraquarks with $J^P=0^+ , 2^+$ are provided.

 In the `type-II' diquark model we propose, diquarks are more resembling compact bosonic building blocks. Indeed we are neglecting spin-spin interactions between different diquarks as we suppose that the size of the entire tetraquark is consistently larger than the size of its building blocks. As for the color force, the diquark-antidiquark pair is described as a bound state of two `point-like'  color sources: the same configuration of a quark-antiquark system. For this reason we make the hypothesis that the spacings in radial excitations could closely resemble those observed in standard $P$-wave charmonia, as indicated by the $Z(4430)-Z(3990)$ mass difference.

In parallel with $Z$ states, there have been extensive experimental investigations on the $Y$ states, pioneered by the Babar discovery of the $Y(4260)$ resonance in $e^+ e^-$ annihilation with Initial State Radiation~\cite{y4260babar}. A considerable number of  $J^{PC}=1^{--}$ states has been observed by Babar and Belle, although several not confirmed. A non exhaustive survey (see also~\cite{brambilla2014} and~\cite{cyuan}) includes: $Y(4008)$ and $Y(4260)$ decaying into $J/\psi$~\cite{ybelle1},\cite{ybabar2},  $Y(4360)$ and $Y(4660)$, decaying into $\psi(2S)$~\cite{ybabar3},\cite{ybelle2}, and $Y(4630)$\cite{ybelle3}, not being clear if the last two are different or are the same particle\footnote{The latter option was proposed in~\cite{cot}.}.  BES III has recently studied decay channels with $h_c(1P)$, which give access to $Y$ states dominated by $c\bar c$ configurations with $S_{c\bar c}=0$, with the possible indication of one narrow state, $Y(4220)$, and  a second wider one, $Y(4290)$ (see~\cite{cyuan,yuanczbes}).

Negative parity states have to be $P$-wave excitations since the basic diquark-antidiquark relative parity is positive.  We note in this paper that four $Y$ states with  $L=1$ are expected, separated by fine and hyperfine mass differences due to spin-orbit and spin-spin couplings, and one with $L=3$. 

Tentatively, we identify $Y(4360)$ and  $Y(4660)$ with the $n=2$ radial excitations of $Y(4008)$ and $Y(4260)$, on the basis of their decay into $\psi(2S)$ and of mass differences very similar to the mass differences of the radial excitations of $\chi_{(c,b)J}(2P)-\chi_{(c,b)J}(1P)$ charmonia.

The selection rule corresponding to $S_{c\bar c}$ conservation leads to identify $Y(4008)$, $Y(4260)$ and $Y(4630)$ with the $n=1, L=1$ states with a dominant component $s_{c\bar c}=1$. The scheme may accomodate also one of the two possible states with dominant decay into $h_c$, indicating a dominant $s_{c\bar c}=0$ component, namely either $Y(4220)$ or $Y(4290)$, but not both. An experimental clarification of the real situation in the $h_c 2\pi$ channel is needed for further progress.

The $Y(4260)$ is assigned the same quark spin structure of the $X(3872)$, making  an electromagnetic, $E1$, transition possible
\begin{equation}
Y(4260) \to \gamma + X(3872)
\label{E1}
\end{equation}
a decay observed by BES III~\cite{Ablikim:2013dyn}. We discuss the selection rules of similar $E1$ transitions of the other $Y$ states, which could provide an effective tool to determine the internal spin structure of $Y$ and $X$ states.

In this paper we do not explore the case of exotic hadrons with hidden charm and strangeness. 
Tetraquarks $[cs][\bar c \bar s]$ states were considered {\it e.g.} in~\cite{dfp} where it was suggested to study the decay channels into $J/\psi \phi $ and $D_s^{(*)}D_s^{(*)}$ . Successively, the first tetraquark candidate decaying into $J/\psi \phi $ was observed by CDF~\cite{y4140cdf}, the $Y(4140)$, recently confirmed by CMS~\cite{y4140}. Its mass and quantum numbers fall within  the spectrum predicted in~\cite{dfp}~\footnote{Expecially a second peak, observed by CDF at $Y(4274)$ in the same decay mode, but not (yet) confirmed by other experiments.}. However, the LHCb collaboration reported a negative result in~\cite{y4140lhcb} and similar results from Belle can be found in~\cite{y4140belle}.

In conclusion, we find that the tetraquark scheme with the new Ansatz on spin-spin couplings is able to reproduce the main features of the spectrum of the observed $X,Y,Z$ particles. This makes even more puzzling the remarkable lack of evidence for a neutral state close to the X(3872) and for its charged counterpart. Further experimental and theoretical investigations are needed, to get to a satisfactory picture of the exotic charmonia.

\subsection*{$S$-wave Tetraquarks}

Tetraquarks in $S$-wave have positive parity and have been classified in~\cite{xmai} according to the diquark and antidiquark spin, $s_{qc}=s$ and $s_{\bar q \bar c}=\bar s$, respectively, and the resulting angular momentum, $J$. For each value of these quantum numbers we have four charge states, with isospin $I=1,0$. Neutral states have a definite charge conjugation, $C=\pm 1$, while we can assign a G-parity, $G=C(-1)^I= -C$ to the charged states, which is conserved in their decays. 

In the  $|s,\bar s\rangle_J$ basis we have the following states
\begin{eqnarray} 
&& {\rm J^{P}}=0^{+}\quad C=+ \quad X_0= |0,0\rangle _0,~X_0^\prime=|1,1\rangle_0\label{zero++} \\
&& {\rm J^{P}}=1^{+}\quad C=+\quad X_1= \frac{1}{\sqrt{2}}\left(|1,0\rangle _1+|0,1\rangle _1\right)\label{uno++} \\
&& {\rm J^{P}}=1^{+}\quad G=+\quad  Z=\frac{1}{\sqrt{2}}\left(|1,0\rangle _1-|0,1\rangle _1\right),~Z^{\prime}=|1,1\rangle_1\label{uno+-} \\
&& {\rm J^{P}}=2^{+}\quad C=+\quad X_2= |1,1\rangle _2 \label{due++} 
\end{eqnarray} 
We use the symbol $Z$ when the charged  states  have been identified and give the corresponding $G$-parity, and use the symbol $X$ for all other cases, reporting the $C$ conjugation of the neutral state.

We identify: $X_1=X(3872)$, while the physical $Z(3900)$ and $Z(4020)$ are identified with the linear combinations of $Z$ and $Z^\prime$ which diagonalize the spin-spin Hamiltonian. 

There are neutral $X$ states quoted in~\cite{brambilla2014,Brambilla:2010cs}, which could be identified with $X_0^\prime$ and $X_2$, notably $X(3915)$ and $X(3940)$, as we discuss below. 

There are indications of neutral counterparts of the $Z$ states~\cite{Xiao:2013iha}, but charged counterparts for the $X$ states have not been detected so far. 

Searches by BaBar~\cite{Aubert:2004zr} and Belle~\cite{Choi:2011fc} exclude the pure $I=1$ assignment of $X(3872)$, however a mixed $I=1$ and $I=0$ seems still possible. The possibility a of a very broad $I=1$ state is considered in~\cite{Terasaki:2011ma}.

It is convenient to put into evidence the heavy quark spin, by introducing the basis where the spins of each quark-antiquark pair are diagonal, which we denote by: $|s_{q\bar q}, s_{c \bar c}\rangle_J$. Charge conjugation is given by~\footnote{The formula holds for states with $L=0$, $L$ being the relative orbital angular momentum of the diquark-antidiquark pair; for general $L$ the formula is $C=(-1)^{L+s_{q\bar q}+s_{c \bar c}}$.}
\begin{equation}
C=(-1)^{s_{q\bar q} + s_{c \bar c}}
\end{equation}
so that states with $C=+1$ (alternatively, $C=-1$ and $G=+1$), have to have equal (unequal) quark-antiquark spins. 

It is not difficult to see that~\footnote{
A spin zero diquark in the color antitriplet channel is defined by $[q_1q_2]_i=\epsilon_{ijk}(\bar q_1^j)_c \gamma_5 q_2^k$, which indeed, apart from a $(-i)$ phase factor, corresponds to the bispinor expression $\epsilon_{ijk}(q_1^j)^T\sigma^2q_2^k$. Therefore, as far as the spin is concerned, the tetraquark state can be described by $(q_1^T\sigma^2 q_2)( \bar q_3^T\sigma^2 \bar q_4) $. Using appropriate normalizations, we can define (using {\it e.g.} $q_1=c,q_2=q,\bar q_3=\bar q,q_4=\bar c$)
\bea
&&|0,0\rangle_0\equiv\frac{1}{2}\sigma^2\otimes \sigma^2  \\
&&|1,1\rangle_0\equiv\frac{1}{2\sqrt{3}}\sigma^2\sigma^i\otimes \sigma^2\sigma^i  
\label{unun}
\eea
where the $i$ index is summed in the latter and the normalization comes from the request: $1/12 \sum_i {\rm Tr} ((\sigma^{i})^{T}\sigma^i)^2=1$. Next we use the completeness relation  $1/2\; \bm \sigma_{ad}\cdot\bm \sigma_{cb}+1/2\; \delta_{ad}\delta_{cb}=\delta_{ab}\delta_{cd}$ which immediately allows to sort out the spin of $c\bar c $ and $q\bar q $  observing that  
$(c_s\sigma^2_{sa}\delta_{ab} q_b)( \bar q_r\sigma^2_{rc} \delta_{cd}\bar c_d) $ contains $\delta_{ab}\delta_{cd}$. Indeed, substituting the completeness relation leads to~(\ref{zero++nb}). Eq.~(\ref{zeroprime++nb}) can  be obtained taking~(\ref{unun}), where we have $(c_s\sigma^2_{sa}\sigma^i_{ab}q_b)(\bar q_r\sigma^2_{rc}\sigma^i_{cd}\bar c_d)$, and using the relation $3/2\; \delta_{ad}\delta_{cb}-1/2\; \bm \sigma_{ad}\cdot\bm \sigma_{cb}=\bm \sigma_{ab}\cdot\bm \sigma_{cd}$.
}
\begin{eqnarray} 
 X_0&=&\frac{1}{2}|0_{q\bar q},0_{c \bar c}\rangle_0+\frac{\sqrt{3}}{2}|1_{q\bar q},1_{c \bar c}\rangle_0,  \label{zero++nb} \\
X_0^\prime&=& \frac{\sqrt{3}}{2}|0_{q\bar q},0_{c \bar c}\rangle_0-\frac{1}{2}|1_{q\bar q},1_{c \bar c}\rangle _0 \label{zeroprime++nb}\\
X_1&=&|1_{q\bar q}, 1_{c \bar c}\rangle_1 \label{uno++nb}\\
 Z&=&\frac{1}{\sqrt{2}}\left(|1_{q\bar q}, 0_{c \bar c}\rangle_1-|0_{q\bar q}, 1_{c \bar c}\rangle_1\right) \\
Z^\prime&=&\frac{1}{\sqrt{2}}\left(|1_{q\bar q}, 0_{c \bar c}\rangle_1+|0_{q\bar q}, 1_{c \bar c}\rangle_1\right) \label{uno+-nb}\\
X_2&=&|1_{q\bar q}, 1_{c \bar c}\rangle_2 \label{due++nb}
\end{eqnarray}

A tentative mass spectrum for the $S$-wave tetraquarks was  derived in~\cite{xmai}, based on an extrapolation of the spin-spin interactions in conventional $S$-wave mesons and baryons.

The resulting couplings turn out to be dominated by the $q\bar q$ coupling, pairing the spin of particles in different diquarks, with a much weaker coupling of the $[qc]$ pair in the same diquark. To a first approximation, the Hamiltonian for this case can  simply  be taken as
\begin{equation}
H\approx 2\kappa_{q\bar q}\;\bm s_q\cdot \bm s_{\bar q}=\kappa_{q\bar q}~ s_{q\bar q}(s_{q\bar q} +1)
\label{ansatz1}
\end{equation}

Mass eigenstates are diagonal in the basis where the $q\bar q$  and the $c\bar c$ have definite spin, the states with $s_{q\bar q}=1$ being heavier. One finds
\begin{equation}
X(3872)\approx X_1=|1_{q\bar q}, 1_{c \bar c}\rangle_1,\quad Z(3900)\approx |1_{q\bar q}, 0_{c \bar c}\rangle_1
\end{equation}
but the other $Z$ would have to be lighter than $3900$~MeV~\cite{zc3900}, in contradiction with BES III finding~\cite{4020}.

In addition, $Z(3900)$ would be made essentially by $s_{c\bar c}=0$~\cite{maiaerice}, at variance with heavy spin conservation and its observed decay into $J/\psi$.
 
 \subsection*{Spin Interactions in Tetraquarks: a New Ansatz}
 
 Rather than trying to enforce our prejudices derived from conventional mesons and baryons, the strength of spin interactions in tetraquarks should be derived from the observed masses of tetraquark candidates. Remarkably, there is a simple approximate Ansatz replacing Eq.~(\ref{ansatz1}) which reproduces the correct spectrum. This consists in taking the dominant spin interactions to be the ones {\it within each diquark}
 \begin{equation}
H\approx 2\kappa_{qc}\; \left(\bm s_q\cdot \bm s_c+  \bm s_{\bar q}\cdot \bm s_{\bar c}\right)= \kappa_{qc}\left[ s(s+1)+\bar s (\bar s +1)-3\right]
\label{ansatz2}
\end{equation} 
  
In this approximation, the mass eigenvectors coincide with the states given in Eqs.~(\ref{zero++})~to (\ref{due++}). We are led to identify
\begin{equation}
X(3872)=X_1,\quad Z(3900)\approx Z,\quad Z(4020)\approx Z^\prime
\end{equation}
with the mass ordering
 \begin{equation}
M(X_1)\approx M(Z),\quad M(Z^\prime)-M(Z)\approx 2 \kappa_{qc}
\label{massnew}
\end{equation}
A value of 
 \begin{equation}
 \kappa_{qc}= 67~{\rm MeV}
\label{kappanew}
\end{equation}
reproduces the two mass differences within less than $20$~MeV. 

The value in (\ref{kappanew}) is considerably larger that $ (\kappa_{qc})_{\bar{\bm3}}= 22$~MeV obtained from the $\Sigma_c-\Lambda_c$ mass difference~\cite{xmai} and may indicate that diquarks in tetraquarks are more compact than diquarks in baryons. 

Considering the other states, in the same approximation we find
\begin{eqnarray}
&& M(X_2)\approx M(X_0^\prime)\approx 4000~{\rm MeV}\label{high}\\
&&M(X_0)\approx 3770~{\rm MeV}\label{low}
\end{eqnarray}
We may wish to identify the first two states with the $X(3940)$ and $X(3916)$, respectively. There is no $X$ state yet identified at masses below the $X(3872)$~\cite{Brambilla:2010cs}.

In this scheme $Z(4430)$ is the first radial excitation of the $Z(3900)$, with a mass difference $Z(4430)-Z(3900)= 593$~MeV, very close to $\psi(2S)-\psi(1S)=589$~MeV.

\begin{figure}[htb!]
\centering
\includegraphics[scale=0.7]{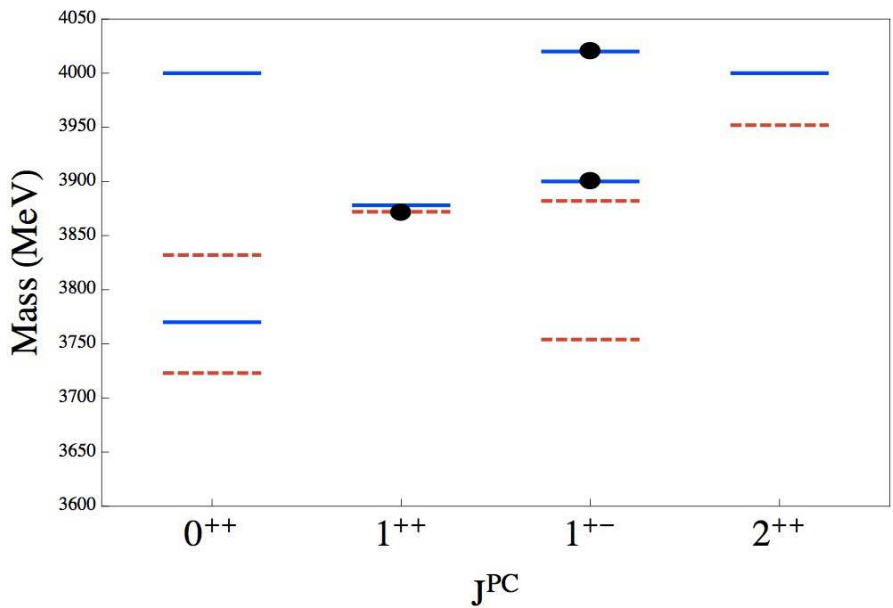}
\caption{\small The dashed (red) levels correspond to the color-spin Hamiltonian introduced in~\cite{xmai}. The solid (blue) levels correspond to the approximation used here~(\ref{ansatz2}). Black disks represent the $X(3872)$, $Z(3900)$ and $Z(4020)$ masses. The $C$ quantum number is the charge conjugation eigenvalue  of the neutral component of the multiplet.  
}
\label{fig1}
\end{figure}

For comparison, we report in Fig.~\ref{fig1} the spectra computed with the old Ansatz~\cite{xmai} and with the new one.

 Finally, we note that the states representing $Z(3900)$ and $Z(4020)$ have now both components of $s_{c\bar c}=1,0$ so as to be at least not in contradiction with the observed  decays~\cite{4020}
 \begin{eqnarray}
&& Z(4020)\to \pi + h_c (1^1P_1),~{\rm seen}\\
&&Z(3900)\to \pi + h_c (1^1P_1),~{\rm seen~(?)} 
 \end{eqnarray}
   
 \subsection*{$Y$ States}

We have previously interpreted the $Y(4260)$ as a tetraquark in $P$-wave, with the composition $[cs][\bar c \bar s]$~\cite{y4260noi} 
due to its dominant decay with $f_0(980)$ production. It was later realized~\cite{hooft,schechter,maiaerice}, that instanton induced effects may mix the $f_0$ with $q\bar q$ states and produce the dominant $f_0\to 2\pi$ decay. The same mechanisms allows $q\bar q \to f_0$ and  a composition for $Y(4260)$ similar to the one of the $S$-wave $X$ and $Z$ states  we are considering.

Tetraquark states with  $J^{PC}(Y)=1^{--}$ can be obtained with odd values of the orbital angular momentum  $L=1, 3$ and diquark and antidiquark spins $s, \bar s=0,1$. We set ourselves in the basis used previously, Eqs.~(\ref{zero++}) to (\ref{due++}), using the notation
\begin{equation}
|s,\bar s; S,L\rangle_{J=1},
\end{equation} 
for the state with total spin $\bm S= \bm s+\bar{\bm s}$ and total angular momentum  $J=1$.

\paragraph{\emph{L=1}.}
We may combine the spin structures in Eqs.~(\ref{zero++}) to (\ref{due++}) with $L=1$ to obtain $J=1$. However, under charge conjugation, the orbital momentum produces a factor $-1$, so we have to keep only spin states classified with $C=+1$. Thus we get the four states
\begin{eqnarray}
&&Y_1=|0,0;0,1\rangle_{1}\label{unoY}\\
&&Y_2=\frac{1}{\sqrt{2}}(|1,0;1,1\rangle_1+ |0,1;1,1\rangle_1\label{dueY}\\
&&Y_3=|1,1;0,1\rangle_{1}\label{treY}\\
&&Y_4= |1,1;2,1\rangle_{1}\label{quattroY}
\end{eqnarray} 

We have ordered the list of $Y$ states according to increasing masses. The orderings $Y_1,Y_2,Y_3,Y_4$ and $Y_1,Y_3,Y_2,Y_4$ correspond to two different particle assignments.   Indeed, inverting $Y_3$ with $Y_2$ has little impact on the choice  of the phenomenological  parameters in the Hamiltonian.  

Comparing with the spin structure of the states in (\ref{zero++nb}) to (\ref{due++nb}), we get the following composition relative to the $c\bar c$ spin
\begin{eqnarray}
&&Y_1:~P(s_{c\bar c}=1):P(s_{c\bar c}=0)=3:1\label{prob1}\\
&&Y_2: ~ P(s_{c\bar c}=1)=1\label{prob2}\\
&&Y_3:~P(s_{c\bar c}=1):P(s_{c\bar c}=0)=1:3\label{prob3}\\
&&Y_4:~P(s_{c\bar c}=1)=1\label{quattroY}
\end{eqnarray}

There is only $Y_3$ expected to decay preferably into the $s_{c\bar c}=0$ state, $h_c(1P)$. 

\paragraph{\emph{L=3}.} There is only one possibility, namely
  \begin{equation}
 Y_5=|1,1;2,3\rangle_{1}
 \label{cinqueY}
\end{equation}
with $S_{c\bar c}=1$.

\paragraph{\emph{Tentative particle assignments.}} As stated in the Introduction, there are indications for more than four states in the region of the $Y(4260)$. Tentatively, we propose the following.
\begin{itemize}

\item We leave aside the $L=3$ state, which is expected to occur at much higher energy (see below);

\item
We interpret the $Y(4360)$ and $Y(4660)$ as radial excitations of $Y(4008)$ and $Y(4260)$, respectively, on the basis of their decay into $\psi(2S)$, analogous to the decay of $Z(4430)$. The relative  mass differences of $350$ and $400$~MeV are in the range of the mass differences for $L=1$ charmonia and bottomonia~\footnote{Mass differences between ground states and radial excited states have been recently analyzed in~\cite{cinax} for tetraquark states and in~\cite{cinay} for charmonia and bottomonia.}
, {\it e.g.} $\chi_{bJ}(2P)-\chi_{bJ}(1P)\approx 360$~MeV whereas $\chi_{cJ}(2P)-\chi_{cJ}(1P)\approx 437$~MeV~\footnote{No data available for $\chi_{c1}(2P)$.}.
 
\item We identfy $Y_{1,2,4}$ with $Y(4008)$, $Y(4260)$ and $Y(4630)$ (decaying into $\Lambda \bar \Lambda$);

\item  We identify $Y_3$ with either  $Y(4290)$ (the broad structure in the $h_c$ channel), or  $Y(4220)$ (the narrow structure)

\end{itemize}

\paragraph{\emph{Spin-orbit and spin-spin interactions.}}  

In the spirit of a first exploration, to confront with the data  we leave aside possible tensor interactions, although we are aware that they play a role in the splitting of $P$-wave charmonia.

We add to the Hamiltonian of $S$-wave tetraquarks an orbital term proportional to $\bm L^2$ and a spin-orbit interaction proportional to $\bm L\cdot\bm S$. 
The restriction of  the spin-spin couplings to the interaction within the same diquark, as discussed before, is more than justified here, due to the angular momentum barrier, and we leave open the possibility that the coupling may take a different value from the $S$-wave case. 

We write
\begin{equation}
M=M_{00}+B_c\frac{\bm L^2}{2} -2a~\bm L\cdot \bm S +2\kappa^\prime_{qc}~\left[(\bm s_q \cdot \bm s_c)+(\bm s_{\bar q} \cdot \bm s_{\bar c})\right]\label{formula1}
\end{equation}
Signs are choosen so that, for $B_c$, $a$, $\kappa$ positive,  energy increases for increasing $\bm L^2$ and $\bm S^2$. With obvious manipulations, we obtain
\begin{eqnarray}
&& M=M_{00}+B_c\frac{L(L+1)}{2} + a~\left[L(L+1)+S(S+1) -2\right] +\notag \\
&&+ \kappa^\prime_{qc}~\left[s(s+1)+\bar s (\bar s +1)-3\right]
\label{formula2}
\end{eqnarray}
namely
\begin{equation}
M=M_{0}+\left(\frac{B_c}{2}+a\right)L(L+1)+a~ S(S+1) + \kappa^\prime_{qc}~[s(s+1)+\bar s (\bar s +1)]
\label{formula3}
\end{equation}
where 
\begin{equation}
M_0=M_{00}-2a - 3 \kappa^\prime_{qc}
\label{m00}
\end{equation}

We then find
\begin{eqnarray}
&&M_1=M_{0}+2\left(\frac{B_c}{2}+a\right)=c \notag \\
&&M_4-M_3=6 a \notag\\
&& M_3-M_2=2\kappa^\prime_{qc}-2a \notag\\
&&M_2-M_1=2\kappa^\prime_{qc}+2a \label{massdiff}
\end{eqnarray}
With four masses and three parameters, we find the relation
\begin{equation}
M_2=\frac{3M_1+M_3+2M_4}{6}
\label{massrel}
\end{equation}
The above formulae require\footnote{So, at variance with the assignment made in~\cite{{y4260noi}}, $Y(4260)$ cannot be identified with $Y_1$.} $M_2>M_1$ and $M_4>M_3$, however the sign of the mass difference $M_3-M_2$ can take either sign, as it is determined by the difference of two constants which are {\it a priori} of a similar size.

We keep fixed the assignment
\begin{equation} 
Y(4260)=Y_2=\frac{1}{\sqrt{2}}(|1,0;1,1\rangle_1+ |0,1;1,1\rangle_1)
\label{4260}
\end{equation} and consider separately the two cases~\footnote{we thank the Referee for suggesting us to consider the two cases on a similar ground. } for $Y_3$
\paragraph{\emph {Y$_3$=Y(4290)}}
The mass relation (\ref{massrel}) is well satisfied by the nominal masses of the $Y$ states, giving
\begin{equation}
(M_2)_{\rm th}= 4262~{\rm MeV}
\end{equation}


We may use the first three equations in~(\ref{massdiff}) to obtain the value of the parameters. Using the nominal $Y$ masses corresponding to the assignment $Y_1=Y(4008)$, $Y_2=Y(4260)$, $Y_3=Y(4290)$ and $Y_4=Y(4630)$ we find (in MeV)
\begin{equation}
a= 56,~\kappa^\prime_{qc}=71
\label{param1}
\end{equation}

\paragraph{\emph {Y$_3$=Y(4220)}}
The mass relation (\ref{massrel}) gives
\begin{equation}
(M_2)_{\rm th}= 4251~{\rm MeV}
\end{equation}
and the value of the parameters are
\begin{equation}
a= 73,~\kappa^\prime_{qc}=53
\label{param2}
\end{equation}


In either  case the value found for $\kappa^\prime_{qc}$ is close to the value in~(\ref{kappanew}), supporting the difference between diquarks in tetraquarks and diquarks in baryons. 

Either structures in the $h_c + 2\pi$ channel can be accommodated in the scheme, but not both. An experimental clarification is needed for further progress.

\paragraph{\emph{The orbital excitation energy.}}
In the new scheme, the spin structure of $Y(4260)$ and $X(3872)$ and their spin interactions are exactly the same and we may obtain the energy of the orbital excitation directly from their mass difference. Starting from Eq.~(\ref{formula2}),  neglecting the difference between $\kappa_{qc}$ and $\kappa^\prime_{qc}$ and using Eq.~(\ref{m00}) we obtain
\begin{eqnarray}
&&M(Y_2)=M(X)+B_c + 2a,~{\it i.e.}\notag \\
&& B_c=278~{\rm MeV}
\label{lexc}
\end{eqnarray}
This value compares well with values found in normal hadrons, see the discussion in~\cite{y4260noi}.  

Finally, a large separation between $Y_5$ and the states $Y_{1-4}$ is implied
\begin{equation}
M_5 = M_2+5B_c+14a\sim 6420~{\rm MeV}
\label{m5}
\end{equation}

\paragraph{\bf The radiative decay of $Y(4260)$.}The identification $Y(4260)=Y_2$ is reinforced by the observation~\cite{Ablikim:2013dyn} of a conspicuous radiative decay mode of $Y(4260)$
\begin{equation}
Y(4260)\to \gamma + X(3872)
\label{raddec}
\end{equation}

The identical spin structure implied in the tetraquark model for the two states suggests this mode to be an unsuppressed $E1$ transition, with $\Delta L=1$ and $\Delta S =0$, similar to the observed transitions of the charmonium $\chi$ states. 

The decay rate of (\ref{raddec}) could provide a first estimate of the radius of the tetraquark.

A comparison of the spin structures in~(\ref{zero++}) to~(\ref{due++}) and~(\ref{unoY}) to~(\ref{quattroY}) provides selection rules for $E1$ transitions between $Y$ and $X$ states that should allow a better identification of the levels, {\it e.g.} wether $Y(4660)$ is or not a radial excitation of  a lower, $P$-wave tetraquark.  With the assignments we made, we expect
\begin{eqnarray}
&&Y_4=Y(4630)\to \gamma + X_2~~(J^{PC}=2^{++})= \gamma + X(3940),~{\rm ??}\\
&&Y_3=Y(4290/4220)\to \gamma + X_0^\prime~~(J^{PC}=0^{++})= \gamma + X(3916),~{\rm ??}\\
&&Y_2=Y(4260)\to \gamma + X_1~~(J^{PC}=1^{++})= \gamma + X(3872),~{\rm seen}\\
&& Y_1=Y(4008)\to \gamma + X_0~~(J^{PC}=0^{++})=\gamma + X(3770~??),~{\rm ??}
\end{eqnarray}

\subsection*{\bf Conclusions and Outlook}

The confirmation of the $Z(4430)$, whose existence has been controversial up to very recently, reinforces the evidence that  hidden-charm tetraquarks exist, as was first predicted  in~\cite{xmai}. Here we choose to use again a diquark-antidiquark representation of the tetraquark but with a new assumption on spin-spin couplings: diquark building blocks are more compact than what was thought before and spin-spin forces outside diquark shells are suppressed. 

This implies a simplified spin-spin interaction Hamiltonian with respect to that postulated in~\cite{xmai}. The new Ansatz allows a good description of the $1^{+-}$ sector, $Z(3900),Z^\prime(4020)$ and $X(3872)$, as described in Fig.~\ref{fig1}.

A consistent description of the $Y$ particles can also be achieved. The $Y(4360)$ and  $Y(4660)$ are identified as the first radial excitations of  $Y(4008)$, $Y(4260)$, respectively. The $L=1$ tetraquarks with predominantly $s_{c\bar c}=1$ are identified with $Y(4008)$, $Y(4260)$ and $Y(4630)$, which decay dominantly in $J/\Psi+2\pi$,  while the fourth state, with dominantly $s_{c\bar c}=0$ component, is identified with either $Y(4290)$ or $Y(4220)$. Only one state is admitted but, at the moment, the two alternatives cannot be distinguished on theoretical grounds.
 
 Finally, the diquark spin structure associated with the $Y(4260)$ accounts for the radiative transition into $X(3872$ observed by BES III as a dominant $E_1$ transition.

The `type-II' diquark-antidiquark model presented here does not yet explain why charged partners of the $X$ and $Y$ states have not been observed. As for the persisting lack of experimental confirmation of two neutral, almost degenerate, $X$ particles at 3872~MeV,  required by the diquark-antidiquark model to account for the strong isospin violating pattern observed in $X$ decays~\cite{xmai}, we believe that this might be due to the sensibility of present experimental analyses. On the other hand this model provides a natural explanation of the quantum numbers of most of the $X,Y,Z$ resonances observed, together with a very reasonable description of their decay rates. 

There are several experimental hints that point to a unified description of $X,Y,Z$ resonances. To quote two striking ones: $i)$ the mass difference between the $Z(4430)$ and $Z(3900)$, the former decaying into $\psi(2S)\;\pi$ and the latter into $J/\psi\;\pi$ $ii)$ the observed radiative transitions between the  $Y(4260)$ and the $X(3872)$. 

After the recent discoveries of the $Z_c(3900)$,  $Z_c(4020)$ and especially $Z(4430)$, we think that the tetraquark option is back with renovated strength and, despite the obvious limitations of the diquark-antidiquark description, we have to observe that it has a descriptive power of the $X,Y,Z$ resonance physics which other explanations cannot provide.

\subsection*{\bf Acknowledgements} 
We would like to thank informative and interesting discussions with A. de Rujula, R. Faccini, B. Gavela, X. Shen, C. Yuan, Q. Zhao. Hospitality by Istituto de Fisica Teorica, Universidad Autonoma de Madrid, is gratefully acknowledged by LM and VR.


\begin{thebibliography}{99}

 \bibitem{z4430lhcb}
 R.~Aaij {\it et al.}  [ LHCb Collaboration],
 arXiv:1404.1903 [hep-ex].
  
\bibitem{z4430belle} 
  S.~K.~Choi {\it et al.}  [BELLE Collaboration],
  Phys.\ Rev.\ Lett.\  {\bf 100}, 142001 (2008)
  [arXiv:0708.1790 [hep-ex]].
 
  \bibitem{z4430belle2}
 K.~Chilikin {\it et al.}  [Belle Collaboration],
 Phys.\ Rev.\ D {\bf 88}, 074026 (2013)
 [arXiv:1306.4894 [hep-ex]].

\bibitem{Aubert:2008aa}
  B.~Aubert {\it et al.}  [BaBar Collaboration],
  Phys.\ Rev.\ D {\bf 79} (2009) 112001
  [arXiv:0811.0564 [hep-ex]].

\bibitem{De Rujula:1976qd}
  A.~De Rujula, H.~Georgi and S.~L.~Glashow,
  Phys.\ Rev.\ Lett.\  {\bf 38} (1977) 317.

\bibitem{cdfprompt}
   A.~Abulencia {\it et al.}  [CDF Collaboration],
 Phys.\ Rev.\ Lett.\  {\bf 98}, 132002 (2007)
 [hep-ex/0612053].

 \bibitem{cmsprompt}
   S.~Chatrchyan {\it et al.}  [CMS Collaboration],
 JHEP {\bf 1304}, 154 (2013)
 [arXiv:1302.3968 [hep-ex]].


\bibitem{bigna}
  C.~Bignamini, B.~Grinstein, F.~Piccinini, A.~D.~Polosa and C.~Sabelli,
  Phys.\ Rev.\ Lett.\  {\bf 103} (2009) 162001
  [arXiv:0906.0882 [hep-ph]];
  C.~Bignamini, B.~Grinstein, F.~Piccinini, A.~D.~Polosa, V.~Riquer and C.~Sabelli,
  Phys.\ Lett.\ B {\bf 684}, 228 (2010)
  [arXiv:0912.5064 [hep-ph]];
 A.~Esposito, F.~Piccinini, A.~Pilloni and A.~D.~Polosa,
  J.\ Mod.\ Phys.\  {\bf 4}, 1569 (2013)
  [arXiv:1305.0527 [hep-ph]].


\bibitem{arto}
P.~Artoisenet and E.~Braaten,
  Phys.\ Rev.\ D {\bf 81} (2010) 114018
  [arXiv:0911.2016 [hep-ph]];
 F.~-K.~Guo, U.~-G.~Mei§ner and W.~Wang,
 Commun.\ Theor.\ Phys.\  {\bf 61}, 354 (2014)
 [arXiv:1308.0193 [hep-ph]];
 F.~-K.~Guo, U.~-G.~Mei§ner and W.~Wang,
 arXiv:1402.6236 [hep-ph];
 F.~-K.~Guo, U.~-G.~Mei§ner, W.~Wang and Z.~Yang,
 arXiv:1403.4032 [hep-ph].

\bibitem{inprep} A.~Guerrieri, A.~Pilloni, F. Piccinini, A.D.~Polosa, {\it in prep.}

\bibitem{Rosner}
  J.~L.~Rosner,
  Phys.\ Rev.\ D {\bf 76} (2007) 114002
  [arXiv:0708.3496 [hep-ph]].

\bibitem{Meng}
  C.~Meng and K.-T.~Chao,
  arXiv:0708.4222 [hep-ph];
  T.~Branz, T.~Gutsche and V.~E.~Lyubovitskij,
  Phys.\ Rev.\ D {\bf 82} (2010) 054025
  [arXiv:1005.3168 [hep-ph]].

 
  \bibitem{Wang:2013cya}
  Q.~Wang, C.~Hanhart and Q.~Zhao,
  Phys.\ Rev.\ Lett.\  {\bf 111} (2013) 132003
  [arXiv:1303.6355 [hep-ph]]; 
  Q.~Wang, C.~Hanhart and Q.~Zhao,
  Phys.\ Lett.\ B {\bf 725} (2013) 1-3,  106
  [arXiv:1305.1997 [hep-ph]].
 
  \bibitem{Qiao}
  C.~-F.~Qiao,
  J.\ Phys.\ G {\bf 35} (2008) 075008
  [arXiv:0709.4066 [hep-ph]].

\bibitem{Bugg}
  D.~V.~Bugg,
  J.\ Phys.\ G {\bf 35} (2008) 075005
  [arXiv:0802.0934 [hep-ph]].

\bibitem{Matsuki}
  T.~Matsuki, T.~Morii and K.~Sudoh,
  Phys.\ Lett.\ B {\bf 669} (2008) 156
  [arXiv:0805.2442 [hep-ph]].
  
\bibitem{Lee}
  S.~H.~Lee, A.~Mihara, F.~S.~Navarra and M.~Nielsen,
  Phys.\ Lett.\ B {\bf 661} (2008) 28
  [arXiv:0710.1029 [hep-ph]].
  
\bibitem{Bracco}
  M.~E.~Bracco, S.~H.~Lee, M.~Nielsen and R.~Rodrigues da Silva,
  Phys.\ Lett.\ B {\bf 671} (2009) 240
  [arXiv:0807.3275 [hep-ph]].


\bibitem{xmai}
  L.~Maiani, F.~Piccinini, A.~D.~Polosa and V.~Riquer,
  Phys.\ Rev.\ D {\bf 71} (2005) 014028
  [hep-ph/0412098].

\bibitem{aali} 
  A.~Ali, C.~Hambrock, I.~Ahmed and M.~J.~Aslam,
  Phys.\ Lett.\ B {\bf 684}, 28 (2010)
  [arXiv:0911.2787 [hep-ph]];
 A.~Ali, C.~Hambrock and W.~Wang,
  Phys.\ Rev.\ D {\bf 85}, 054011 (2012)
  [arXiv:1110.1333 [hep-ph]].

\bibitem{z4430} 
  L.~Maiani, A.~D.~Polosa and V.~Riquer,
  arXiv:0708.3997 [hep-ph].

 \bibitem{Ablikim:2013mio}
  M.~Ablikim {\it et al.}  [BESIII Collaboration],
  Phys.\ Rev.\ Lett.\  {\bf 110} (2013) 252001. 
  
  \bibitem{Liu:2013dau}
  Z.~Q.~Liu {\it et al.}  [Belle Collaboration],
  Phys.\ Rev.\ Lett.\  {\bf 110} (2013) 252002
  [arXiv:1304.0121 [hep-ex]].
 
  \bibitem{Xiao:2013iha}
 T.~Xiao, S.~Dobbs, A.~Tomaradze and K.~K.~Seth,
 Phys.\ Lett.\ B {\bf 727}, 366 (2013)
 [arXiv:1304.3036 [hep-ex]].

  \bibitem{4020} M.~Ablikim {\it et al.}  [BESIII Collaboration],
 Phys.\ Rev.\ Lett.\  {\bf 111}, 242001 (2013)
 [arXiv:1309.1896 [hep-ex]].
    
\bibitem{Ablikim:2013emm}
 M.~Ablikim {\it et al.}  [BESIII Collaboration],
 Phys.\ Rev.\ Lett.\  {\bf 112}, 132001 (2014)
 [arXiv:1308.2760 [hep-ex]];
  
 \bibitem{noirassegna}
   N.~Drenska, R.~Faccini, F.~Piccinini, A.~Polosa, F.~Renga and C.~Sabelli,
  Riv.\ Nuovo Cim.\  {\bf 033}, 633 (2010)
  [arXiv:1006.2741 [hep-ph]].

 \bibitem{maiaerice} L. Maiani, talk given at the Erice School of Subnuclear Physic, June 30, 2013, EMFCSC, Erice, Italy, July 2013,  [arXiv:1404.6618 [hep-ph]].
 
 \bibitem{y4260babar}
 B.~Aubert {\it et al.}  [BaBar Collaboration],
  Phys.\ Rev.\ Lett.\  {\bf 95}, 142001 (2005)
  [hep-ex/0506081].
 
  \bibitem{brambilla2014} N. Brambilla {\it et al.}, arXiv:1404.3723v1 [hep-ph].
  
 \bibitem{cyuan} C. Yuan, arXiv: 1404.7768 [hep-ph].

\bibitem{ybelle1} C. Z. Yuan {\it et al.}  [Belle Collaboration], Phys.\ Rev.\ Lett.\  {\bf 99}, 182004 (2007), arXiv:0707.2541v2 [hep-ex]; Z. Q. Liu {\it et al.}  [Belle Collaboration], Phys.\ Rev.\ Lett.\  {\bf 110}, 252002 (2013), arXiv:1304.0121v2 [hep-ex]

\bibitem{ybabar2} J. P. Lees {\it et al.}  [BaBar Collaboration], Phys.\ Rev. {\bf D86}, 051102 (2012), arXiv:1204.2158v1 [hep-ex]. 

\bibitem{ybabar3} B. Aubert {\it et al.}  [BaBar Collaboration], Phys.\ Rev.\ Lett.\  {\bf 98} 212001 (2006), arXiv:hep-ex/0610057v2; J. P. Lees {\it et al.}  [BaBar Collaboration],  arXiv:1211.6271v2 [hep-ex].

\bibitem{ybelle2} X. L. Wang {\it et al.}  [Belle Collaboration], Phys.\ Rev.\ Lett.\  {\bf 99}, 142002 (2007), arXiv:0707.3699v2 [hep-ex].

\bibitem{ybelle3} G. Pakhlova {\it et al.}  [Belle Collaboration], Phys.\ Rev. \ Lett. {\bf 101}, 172001 (2008), arXiv:0807.4458v2 [hep-ex].

  \bibitem{cot} 
  G.~Cotugno, R.~Faccini, A.~D.~Polosa and C.~Sabelli,
  Phys.\ Rev.\ Lett.\  {\bf 104}, 132005 (2010)
  [arXiv:0911.2178 [hep-ph]].
  
  \bibitem{yuanczbes} C.~Z.~Yuan, Chinese Physics {\bf C38}, 043001 (2014). 

  \bibitem{Ablikim:2013dyn}
  M.~Ablikim {\it et al.}  [BESIII Collaboration],
  arXiv:1310.4101 [hep-ex].

\bibitem{dfp} 
  N.~V.~Drenska, R.~Faccini and A.~D.~Polosa,
  Phys.\ Rev.\ D {\bf 79}, 077502 (2009)
  [arXiv:0902.2803 [hep-ph]].

 \bibitem{y4140cdf} 
   T.~Aaltonen {\it et al.}  [CDF Collaboration],
 arXiv:1101.6058 [hep-ex].

\bibitem{y4140}
   S.~Chatrchyan {\it et al.}  [CMS Collaboration],
 arXiv:1309.6920 [hep-ex].
 
 \bibitem{y4140lhcb} 
  RAaij {\it et al.}  [LHCb Collaboration],
  Phys.\ Rev.\ D {\bf 85}, 091103 (2012)
  [arXiv:1202.5087 [hep-ex]].

\bibitem{y4140belle} 
  C.~P.~Shen {\it et al.}  [Belle Collaboration],
  Phys.\ Rev.\ Lett.\  {\bf 104}, 112004 (2010)
  [arXiv:0912.2383 [hep-ex]].

   \bibitem{Brambilla:2010cs}
  N.~Brambilla, S.~Eidelman, B.~K.~Heltsley, R.~Vogt, G.~T.~Bodwin, E.~Eichten, A.~D.~Frawley and A.~B.~Meyer {\it et al.},
  Eur.\ Phys.\ J.\ C {\bf 71} (2011) 1534
  [arXiv:1010.5827 [hep-ph]].

\bibitem{Aubert:2004zr}
  B.~Aubert {\it et al.}  [BaBar Collaboration],
  Phys.\ Rev.\ D {\bf 71} (2005) 031501
  [hep-ex/0412051].

\bibitem{Choi:2011fc}
  S.~-K.~Choi, S.~L.~Olsen, K.~Trabelsi, I.~Adachi, H.~Aihara, K.~Arinstein, D.~M.~Asner and T.~Aushev {\it et al.},
  Phys.\ Rev.\ D {\bf 84} (2011) 052004
  [arXiv:1107.0163 [hep-ex]].

\bibitem{Terasaki:2011ma}
  K.~Terasaki,
  Prog.\ Theor.\ Phys.\  {\bf 127} (2012) 577
  [arXiv:1107.5868 [hep-ph]].

  
  \bibitem{zc3900} 
  L.~Maiani, V.~Riquer, R.~Faccini, F.~Piccinini, A.~Pilloni and A.~D.~Polosa,
  Phys.\ Rev.\ D {\bf 87}, no. 11, 111102 (2013)
  [arXiv:1303.6857 [hep-ph]].
  
 
  \bibitem{y4260noi}
L.~Maiani, V.~Riquer, F.~Piccinini and A.~D.~Polosa,
  Phys.\ Rev.\ D {\bf 72}, 031502 (2005)
  [hep-ph/0507062].
  
\bibitem{hooft} 
  G.~'t Hooft, G.~Isidori, L.~Maiani, A.~D.~Polosa and V.~Riquer,
  Phys.\ Lett.\ B {\bf 662}, 424 (2008)
  [arXiv:0801.2288 [hep-ph]].

\bibitem{schechter} A.~H.~Fariborz, R.~Jora and J.~Schechter,
  Phys.\ Rev.\ D {\bf 77} (2008) 094004
  [arXiv:0801.2552 [hep-ph]]. 
  
  \bibitem{cinax} 
  Z.~-G.~Wang,
  arXiv:1405.3581 [hep-ph].
  
  \bibitem{cinay} 
  L.~-P.~He, D.~-Y.~Chen, X.~Liu and T.~Matsuki,
  arXiv:1405.3831 [hep-ph].
  
  
 \end{thebibliography}
\end{document}